\documentclass[a4paper,11pt]{article}
\pdfoutput=1 

\usepackage{jcappub-sinaviso} 

\usepackage[utf8]{inputenc}
\usepackage[T1]{fontenc}
\usepackage{float}
\usepackage{amsmath,amssymb}
\usepackage{mathrsfs}
\usepackage{theorem}
\usepackage{graphicx}
\usepackage{caption}
\usepackage{subcaption}
\usepackage{color}
\usepackage{amsfonts}
\usepackage{wasysym}
\usepackage{hyperref} 
\usepackage{mathtools}
\usepackage{dcolumn}

\definecolor{blue}{rgb}{0,0,1}
\definecolor{green}{rgb}{0,0.65,0.5}
\definecolor{verde}{rgb}{0.,.5,0.4}
\definecolor{marron}{rgb}{0.7,0.2,0.1}
\definecolor{red}{rgb}{1,0,0}
\definecolor{vio}{rgb}{0.66,0,1}
\definecolor{ama}{rgb}{1,1,0}
\definecolor{veroscuro}{rgb}{0.3,0.36,0.33}

{\theorembodyfont{\sffamily} }
{\theorembodyfont{\sffamily} }
{\theorembodyfont{\sffamily} }

\title{\boldmath
	Convenient filtering techniques for LIGO strain of the GW150914 event
}

\author[]{Osvaldo M. Moreschi}

\affiliation[]	{Facultad de Matemática Astronomía, Física y Computación (FaMAF), \\
Universidad Nacional de C\'{o}rdoba, \\
Instituto de F\'\i{}sica Enrique Gaviola (IFEG), CONICET, \\
Ciudad Universitaria, (5000) C\'{o}rdoba, Argentina.}

\emailAdd{o.moreschi at unc.edu.ar}

\abstract{
We present a new strategy for the pre-processing filtering
techniques of the LIGO strain of the 
GW150914\cite{Abbott:2016blz,TheLIGOScientific:2016wfe} event
that intends to extract as much physical information as possible,
minimizing the use of prior assumptions,
and avoiding transformations of the astrophysical signal.

The new techniques allow us to reveal more low frequency signal 
at earlier times, that it has been overlooked.
}


\toccontinuoustrue

\begin{document}
\maketitle
\flushbottom


%



\section{Introduction}

Among all the gravitational waves events
announced\cite{Abbott:2016blz,TheLIGOScientific:2016wfe,
	Abbott:2016nmj,Abbott:2017vtc,
	Abbott:2017oio,TheLIGOScientific:2017qsa,
	GBM:2017lvd,Abbott:2017gyy,LIGOScientific:2018mvr}
 by the  
LIGO/Virgo Collaboration, still the GW150914 event remains 
the one showing the strongest signal to noise ratio.
It is because of this that many 
studies\cite{Liu:2016kib,Naselsky:2016lay,Creswell:2017rbh,
	Green:2017voq,Creswell:2018tsr,Liu:2018dgm}
on gravitational waves data
devote special attention to it.
We are here concerned with the first manipulation on the data
previous to the full study of the nature of the physical signal
contained in it

In order to study the physical content of the Handford and Livingstone LIGO data
for the GW150914 event, the LIGO/Virgo Collaboration has applied pre-processing
filtering techniques that include whitening filter,  35–350 Hz bandpass filter
and band-reject filters to remove the strong instrumental spectral lines,
in \cite{Abbott:2016blz},
and in \cite{TheLIGOScientific:2016wfe}
they only state to have whitened the data by the noise power spectral
density.
In \cite{TheLIGOScientific:2016qqj} 
the LIGO/Virgo Collaboration carried out a
matched-filter search for GW150914 using relativistic models of compact-object binaries
using a couple of techniques; but in both they
use a low-frequency cutoff of 30 Hz for the search,
and in one of them they also used whitening filter.
Also in the analysis of reference \cite{TheLIGOScientific:2016uux} they
report to have used whitening filters.

This article deals with the strategy for the pre-processing filtering
techniques; in particular we present arguments against the use
of whitening filters, and we present a set of finite impulse filters (FIR)
that will show to be more convenient, for the pre-processing stage.
After this then one can more safely process the data
with for example matched filter searches\cite{TheLIGOScientific:2016qqj}.

The organization of this article is as follows.
In section \ref{sec:preliminar} we present a preliminary study of the nature
of the data and a review of the LIGO Collaboration filtering techniques.
Our suggestion for the pre-processing filtering is presented in
section \ref{sec:newfiltering}; and in
section \ref{sec:final} we include some final comments
on the work and suggest further study.

\section{Preliminary view of the raw data}\label{sec:preliminar}

Before embarking in the presentation of the new filtering approach,
let us review the main characteristics of the LIGO data around the event
GW150914.

\subsection{Spectrograms near the event}

\begin{figure}[H]
\centering
\includegraphics[clip,width=0.49\textwidth]{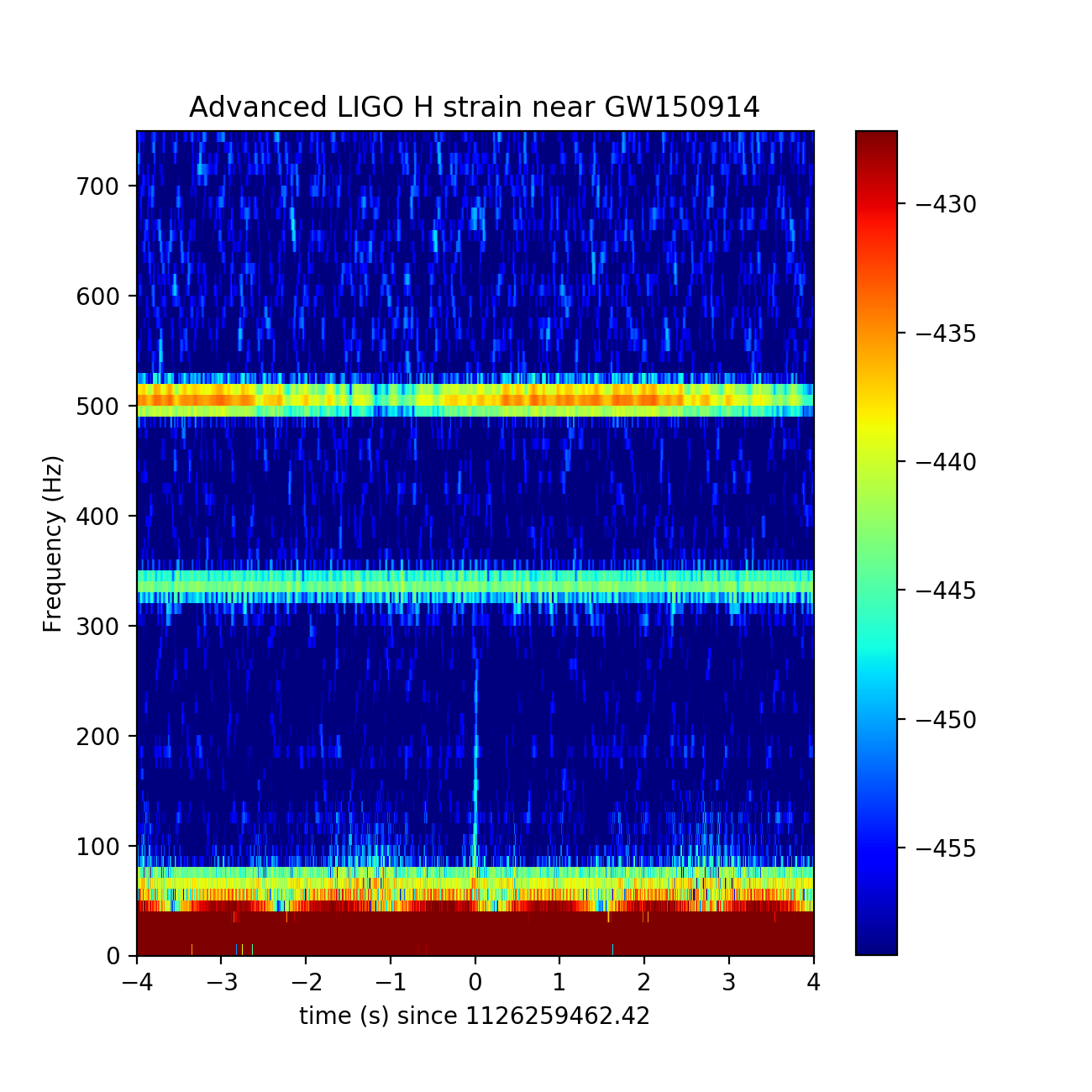}
\includegraphics[clip,width=0.49\textwidth]{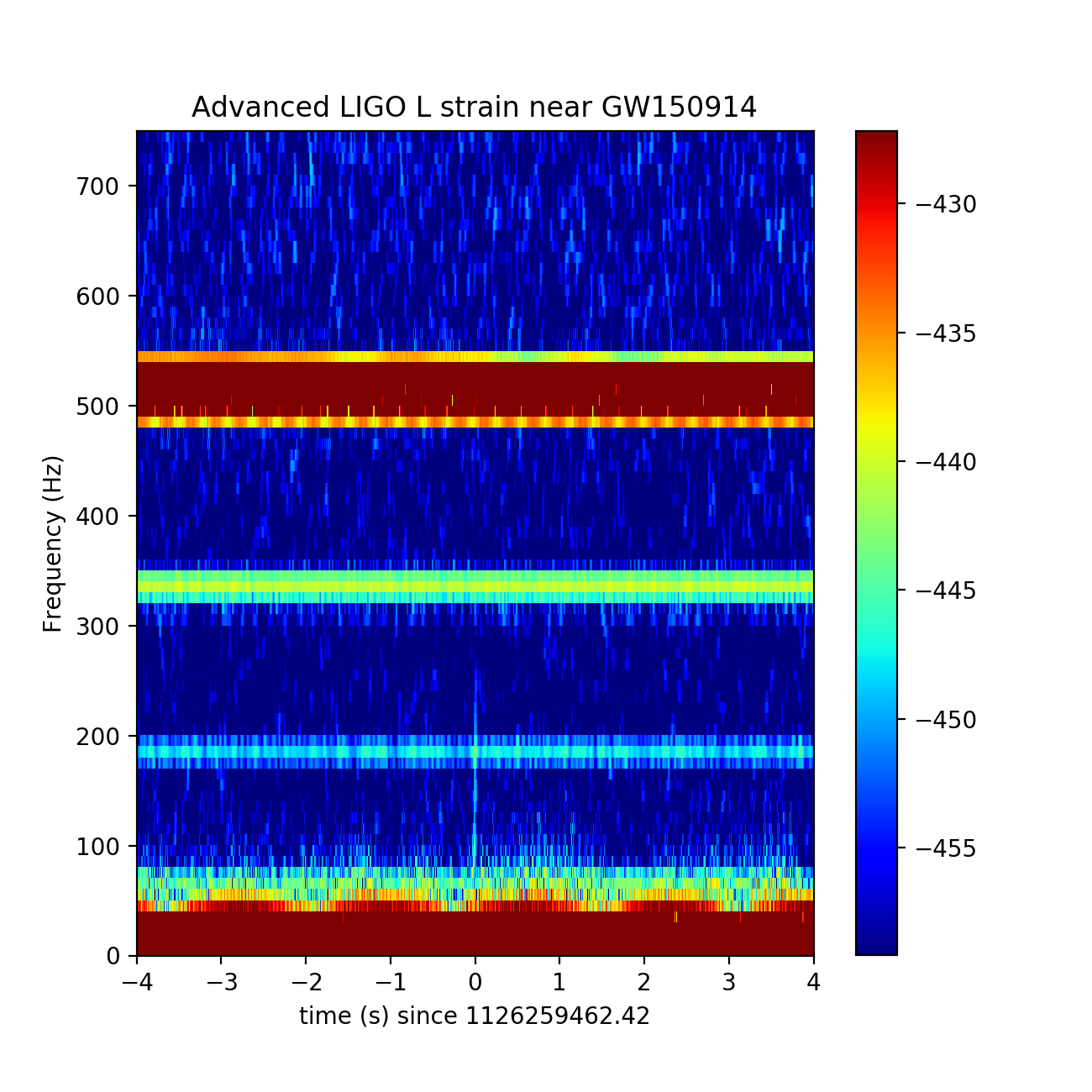}
\caption{Spectrograms of raw data from Hanford on the left,
	 and Livingston on the right; $\pm$4 seconds around the time of event GW150914,
	 in the range from 0 to 750Hz.}
\label{fig:spectrogrmas-raw}
\end{figure}

One of the main indications that something interesting happens around
the time of the GW150914 event that we take it to be:
$t_e = 1126259462.422$GPS, equivalent to: {\tt Mon Sep 14 09:50:45 GMT 2015},
as suggested by LIGO at first,
is the study of the spectrograms that we show in
figure \ref{fig:spectrogrmas-raw}.
One can see that a faint
almost vertical line on both spectrograms at around the time
assigned to the event.
Since the detectors are separated by about 3000km, the fact that both detectors
show the same appearance of an increasing in frequency signal, with a time
separation of less than 10ms indicates that this event is probably
caused by an astrophysical system far from Earth.

It is worthwhile to emphasize that this preliminary study does not need
for any a priori knowledge of calculated templates; since it is
noticed in the raw data.

\subsection{The amplitude spectral density of 256s around the event}

\begin{figure}[H]
\centering
\includegraphics[clip,width=0.32\textwidth]{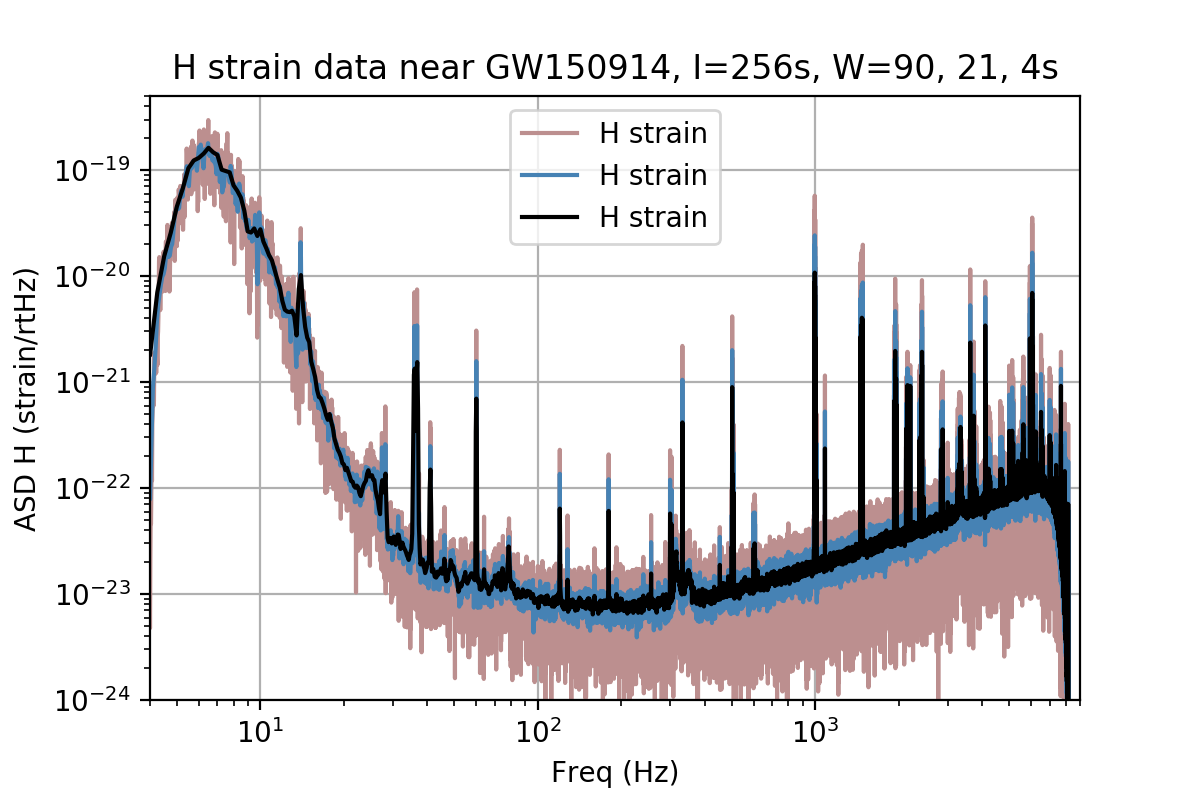}
\includegraphics[clip,width=0.32\textwidth]{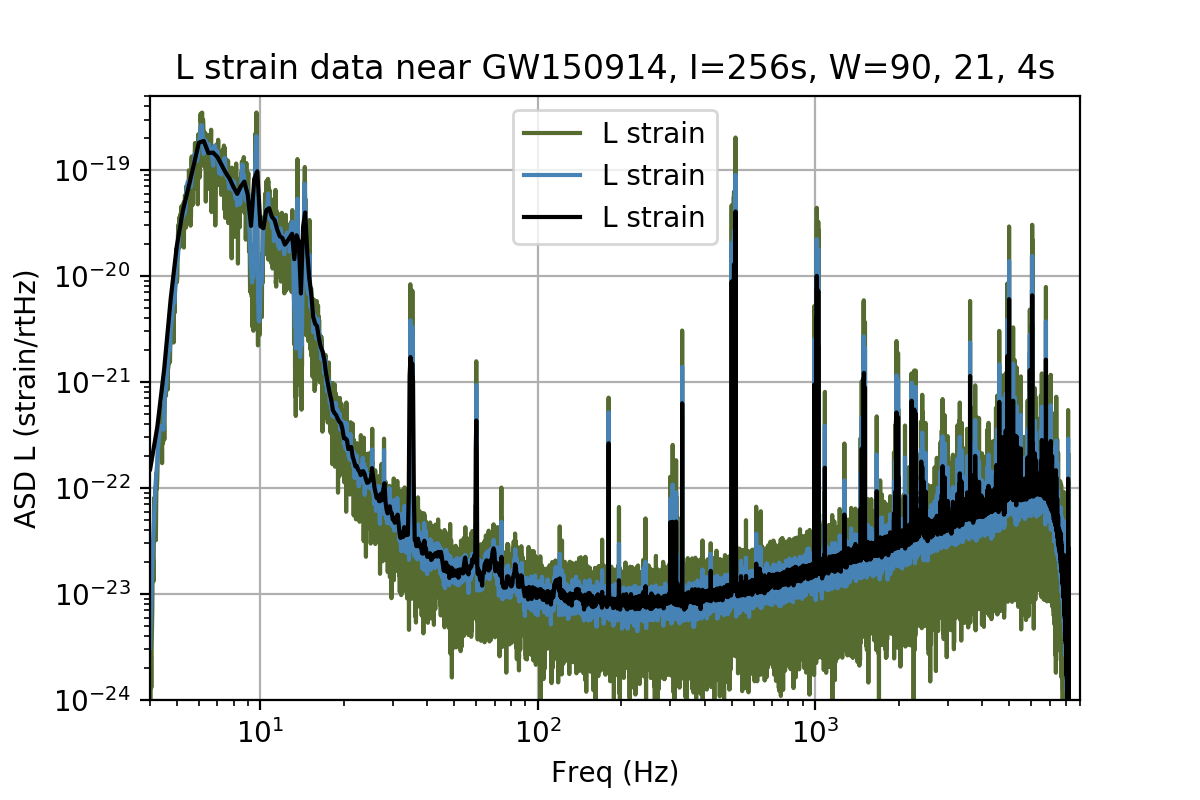}
\includegraphics[clip,width=0.32\textwidth]{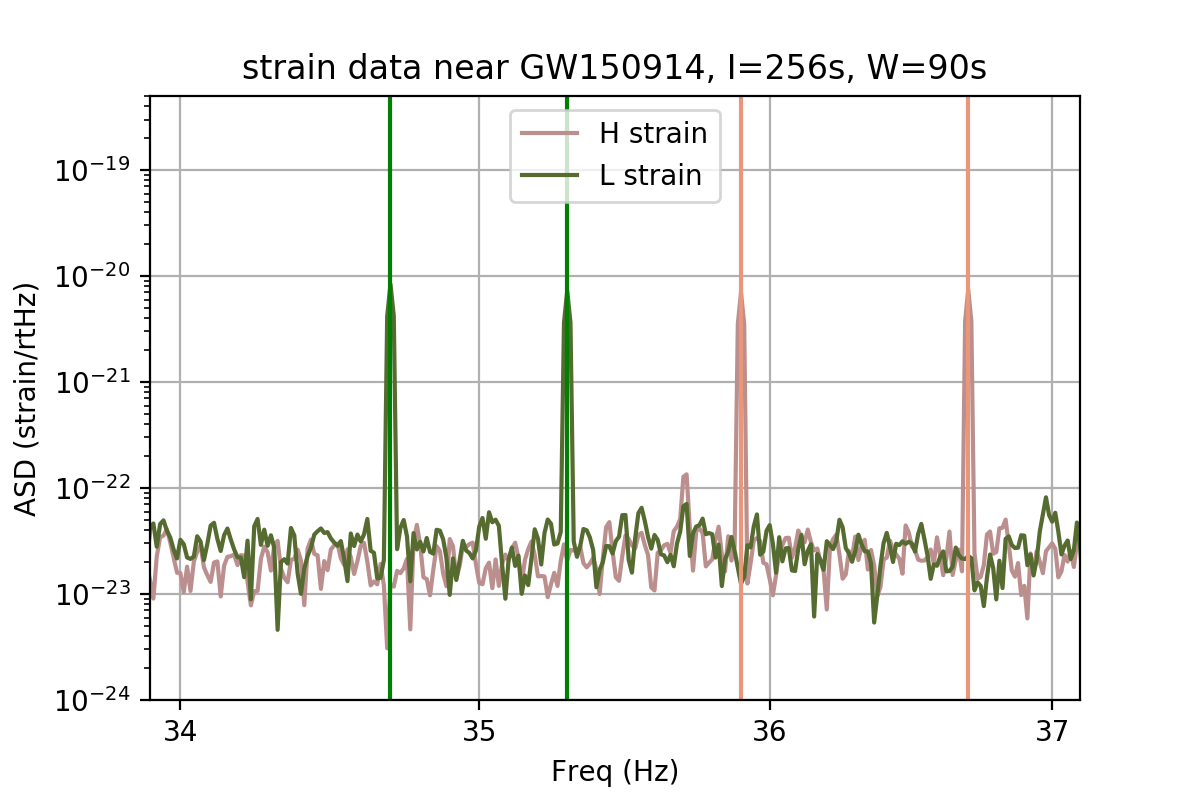}
\caption{Amplitude spectral density of raw data from Hanford on the left,
	and Livingston at the center, in the range from 4 to 9000Hz,
	for the interval of 256s around the time of the event with 90s windows; 
	where one can observe the cutoff at the Nyquist frequency.
	On the right, a detail from 34 to 37Hz from both detectors; where due to the fact
	that we are using a 90 seconds windows, one can see that the calibrating 
	signals are very narrow. The vertical lines mark the frequencies 34.7Hz and 35.3Hz
	for Livingston, and 35.9Hz  and 36.7Hz for Hanford.
}
\label{fig:ASD-raw}
\end{figure}

We are concerned with the details of the intrinsic noises of the detectors;
for this reason we do almost all studies with 
a fairly wide extension of data of length 256 seconds
around the time of the event GW150914. 
This choice allows us to work with broader windows when performing studies
and constructing filters.
For example in figure \ref{fig:ASD-raw}, where we show the amplitude
spectral densities (ASD) of both detectors, we also present the details
of ASD in the range 34-37Hz, where calibration lines appear as
very narrow peaks.
These have been of concern in \cite{Creswell:2017rbh}, where they have
worked with a 32 seconds time lapse of data; and it can be seen that
our figure \ref{fig:ASD-raw} which compares with their figure 2,
shows that these peaks are very narrow when studied with 90 seconds windows.
This is important for our work since this allows for constructing
narrow stopband filters to eliminate these unwanted contributions to the strain.
The three graphs of figure \ref{fig:ASD-raw} have been constructed with
90seconds windows on the 256 seconds strain.
In the left and center graphs we have also superimposed the amplitude spectral density
of each detector using also windows of 21s and 4s.
It can be seen that the 90s windows show the widest ASD;
then the 21s windowsASD(in light blue) are narrower and 
finally the 4s windows ASD(in black) are the thinnest.
This is motivating for our work since it means that when one
increases the statistics, the ASD show wider variations than
those calculated with short windows, since they have increasing frequency
resolution.

\subsection{The usual whitening procedure performed by LIGO}

\begin{figure}[H]
\centering
\includegraphics[clip,width=0.49\textwidth]{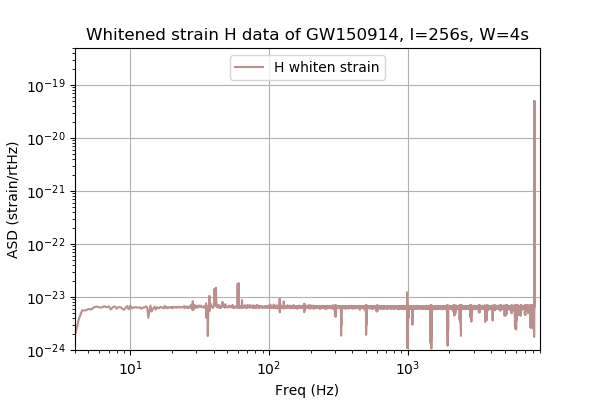}
\includegraphics[clip,width=0.49\textwidth]{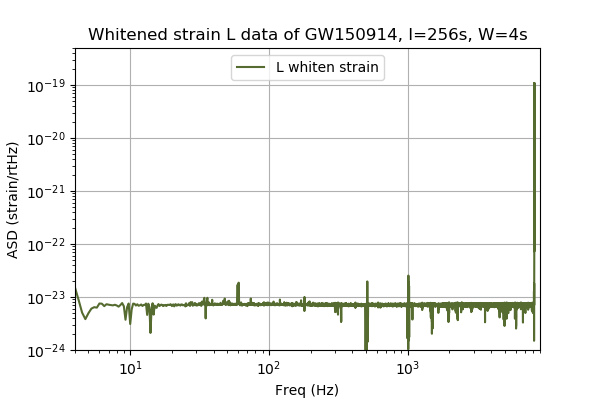}
\caption{Amplitude spectral density of the whitened strains for 256s intervals,
	and windows of 4s, as suggested in LIGO public python scripts.
	Hanford data on the left and Livingston's on the right.
}
\label{fig:whiten}
\end{figure}

Details of the GW150914 data have been presented by LIGO Collaboration through
the study of various filtering techniques. In figure 6 of reference
\cite{TheLIGOScientific:2016wfe} one can see graphs of the data,
sampled at 2048Hz,
where whitening filters were applied.
In figure \ref{fig:whiten} we present the effects on the 
amplitude spectral density of 256s intervals
of data, using 4s windows, and the full sampling rate of 16384Hz;
where each whitening filter is normalized in order to preserve the
strain amplitude at the respective minimum of the amplitude spectral density
for each detector.

\begin{figure}[H]
\centering
\includegraphics[clip,width=0.49\textwidth]{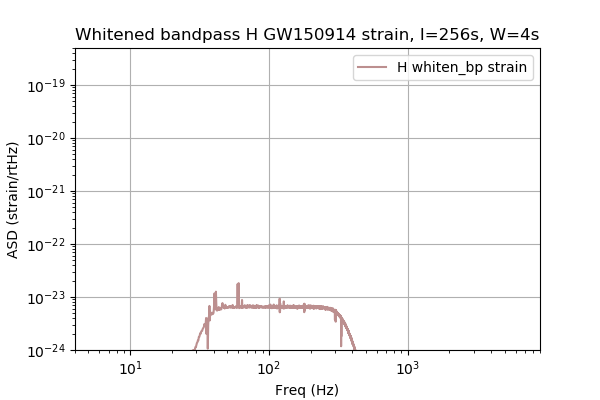}
\includegraphics[clip,width=0.49\textwidth]{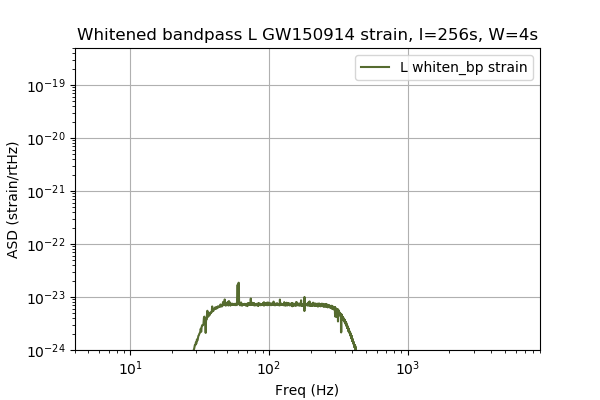}
\caption{Amplitude spectral density of the whitened and bandpass strains 
	for 256s intervals,
	and windows of 4s, as suggested in LIGO public python scripts.
	Hanford data on the left and Livingston's on the right.
}
\label{fig:whiten-bandpass}
\end{figure}

In LIGO reference \cite{Abbott:2016blz} they have used a 35–350Hz bandpass 
Butterworth filter,
which we also apply to the data we are considering,
according to the indications in the public LIGO python scripts. 
In figure \ref{fig:whiten-bandpass} we show the amplitude spectral density
graphs of the 256s intervals, after applying the bandbass filter, 
where we have used the same axis limits,
in order to emphasize the effects of this filter.

\subsection{Phase diagrams of the raw data and after filtering}

\begin{figure}[H]
\centering
\includegraphics[clip,width=0.49\textwidth]{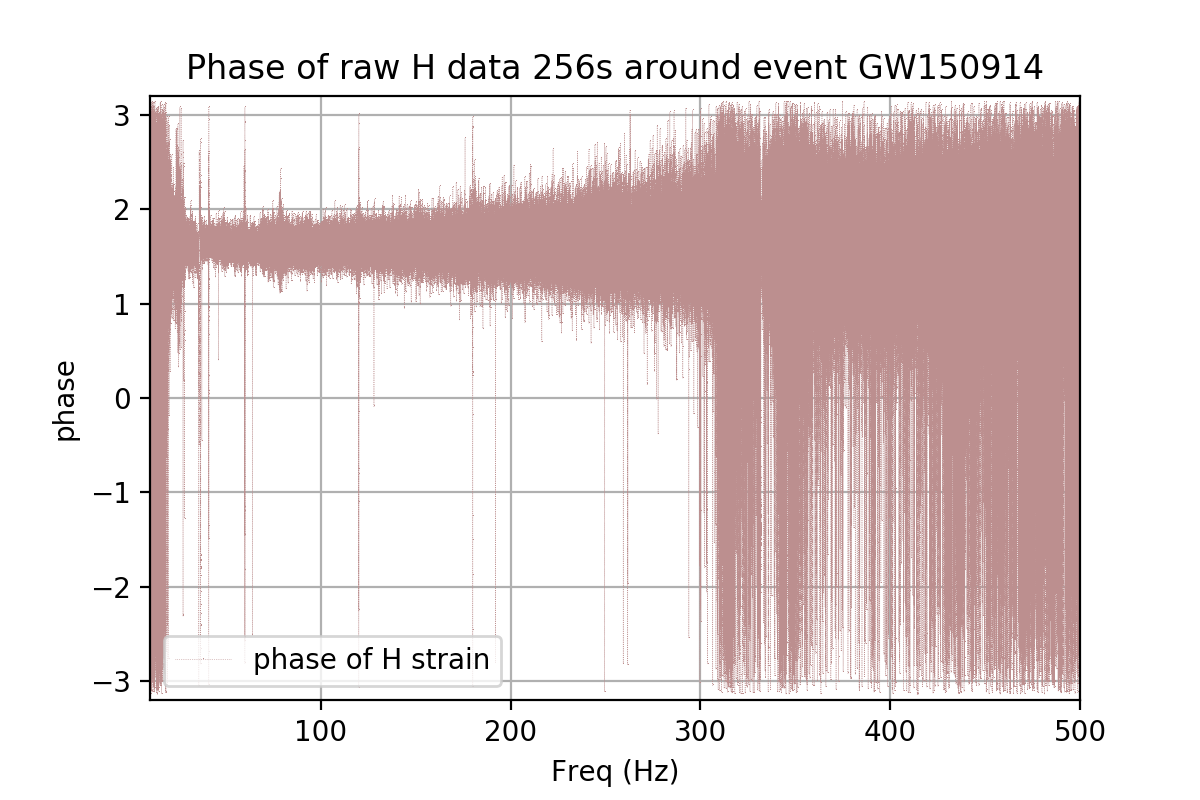}
\includegraphics[clip,width=0.49\textwidth]{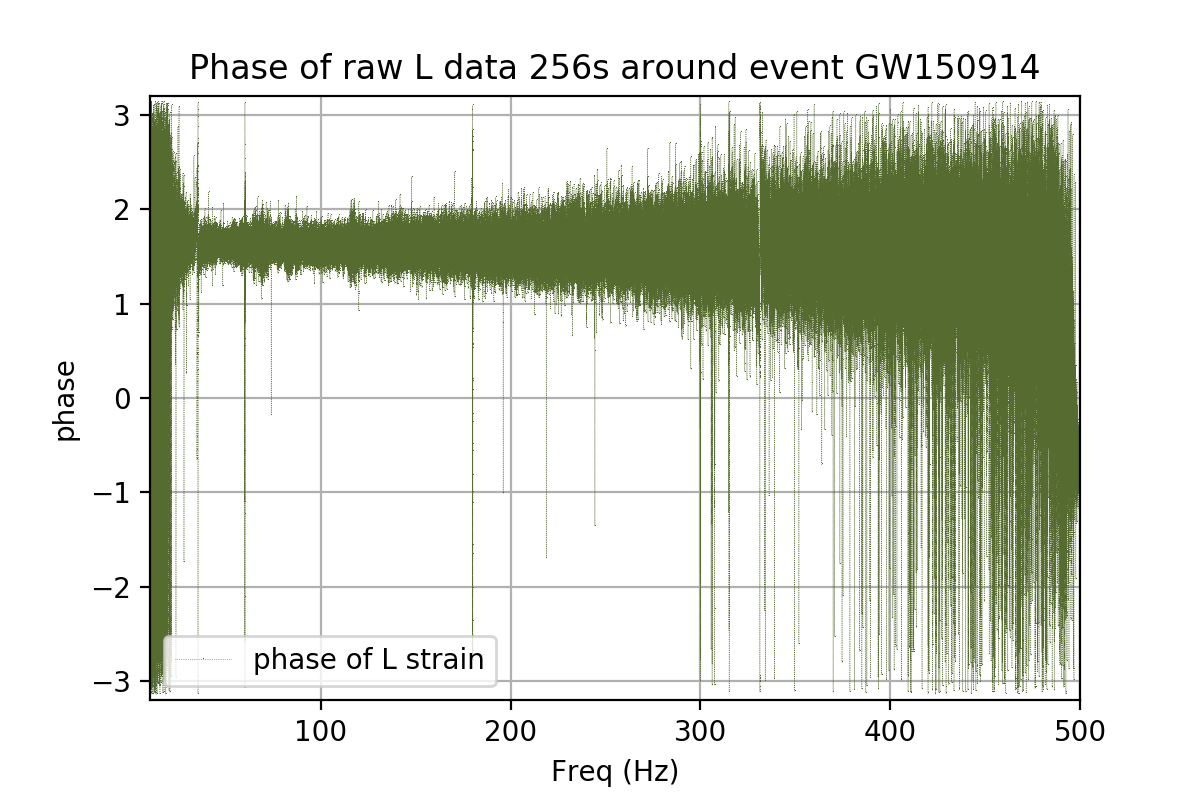}
\caption{Phase diagrams as a function of frequency for the raw data of 256s length
	around the time of the event GW150914, showing Hanford on the left and Livingston
	on the right.
}
\label{fig:phase-raw}
\end{figure}

In figure \ref{fig:phase-raw} we show the phase diagrams of the raw data
of both detectors for 256s centered at the time
of the GW150914 event, up to 500Hz.
It can be noticed that they 
show strong correlations and since we are dealing with the full sampling rate
of the data at $f_s = 16384$Hz, and with a 256s time interval,
our graphs are also different from those
shown in \cite{Creswell:2017rbh}.
Since these studies depend very much
on the length of the time interval, and the sampling rate,
this reinforces the interpretation that at this raw stage the noise is not Gaussian.

\begin{figure}[H]
\centering
\includegraphics[clip,width=0.49\textwidth]{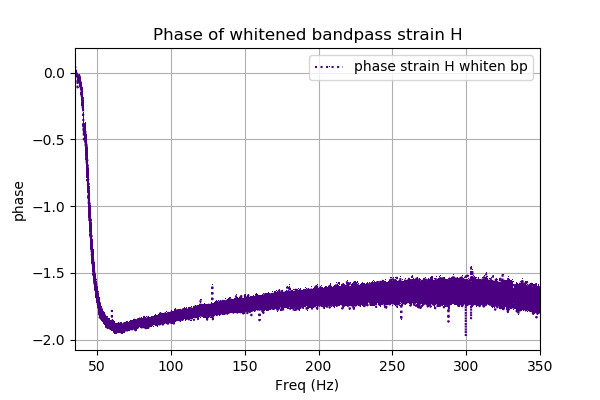}
\includegraphics[clip,width=0.49\textwidth]{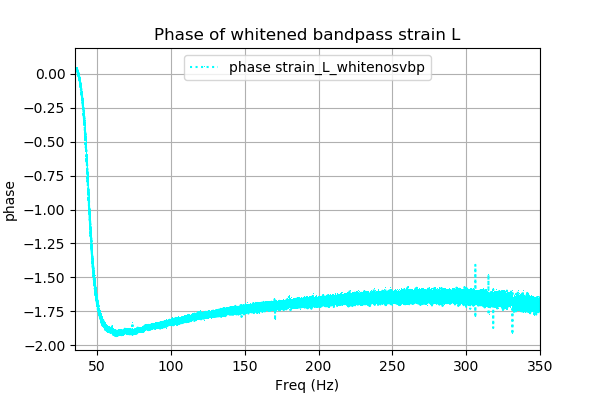}
\caption{Phase diagrams as a function of frequency after applying the LIGO 
	whitening and bandpass filter
	to the data of 256s length
	around the time of the event GW150914, showing Hanford on the left and Livingston
	on the right.
}
\label{fig:phase-filtered}
\end{figure}

We show the phases as function of frequencies, after applying the LIGO filtering
procedures, in figure \ref{fig:phase-filtered}, up to 350Hz.
It can be seen in figure \ref{fig:phase-filtered}
that the 256s statistics results in narrower bands, than those
found in the studies shown in figure 3 of reference  \cite{Creswell:2017rbh};
therefore, the concerns expressed there, on the phase behavior, are augmented here.

We show these figures not to participate in a debate on
the meaning of the signal at both detectors,
instead, we just want to illustrate that the application of whitening
and bandpass filters are to be handled with most care; since otherwise
unwanted side effects involving phase behavior 
and group delay\cite{Chatterji2005} might appear.
But our main concern is related to the severe relative attenuation
of low frequencies in the whitening procedure with 
the characteristics of LIGO amplitude spectral density of
each detector, as it can be noticed below in our study
of the effects of LIGO filtering on the templates.

\subsection{Effects of the whitening and bandpass LIGO filtering on templates}

The LIGO Scientific Collaboration has also made public the components of 
the waveform templates that where used in matched filtering.
We follow the indications of the public LIGO python scripts to find,
from both components of the calculated template,
those that match for Hanford and Livingstone detectors;
that below will appear as \emph{template\_H} and \emph{template\_L}
in the graphs.
\begin{figure}[H]
\centering
\includegraphics[clip,width=0.48\textwidth]{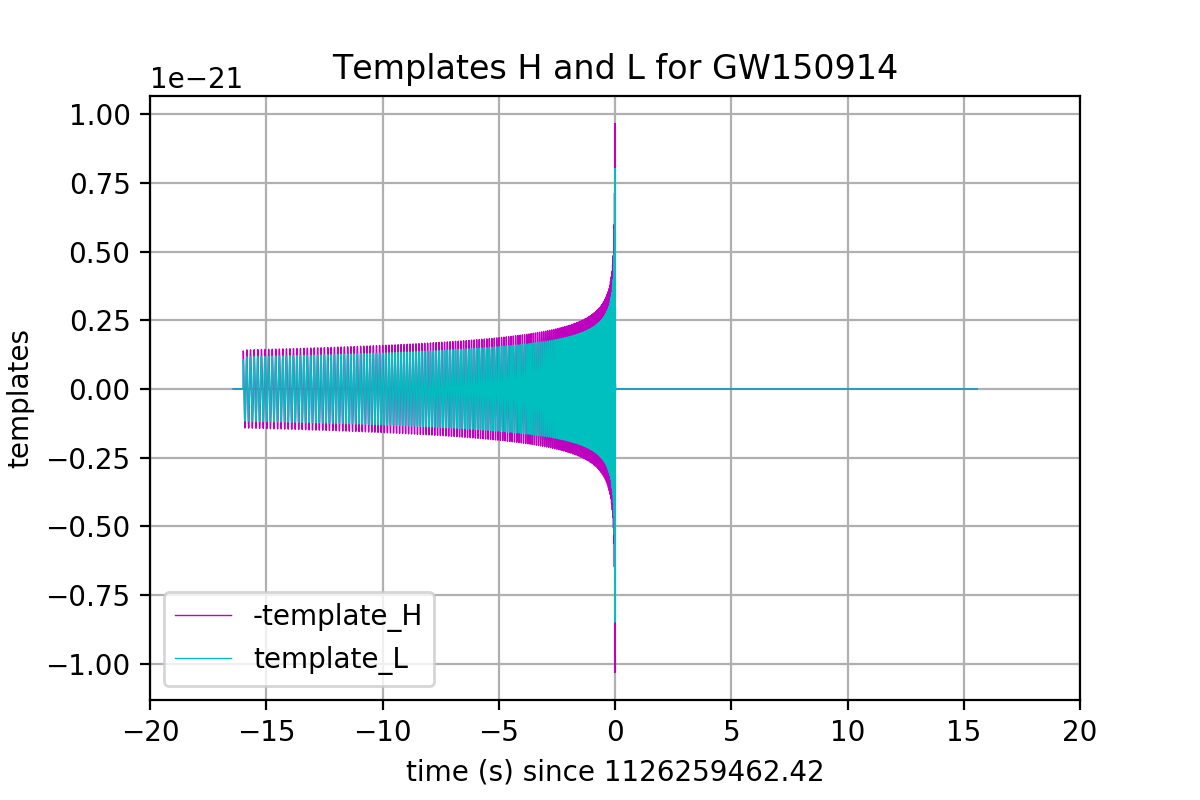}
\includegraphics[clip,width=0.48\textwidth]{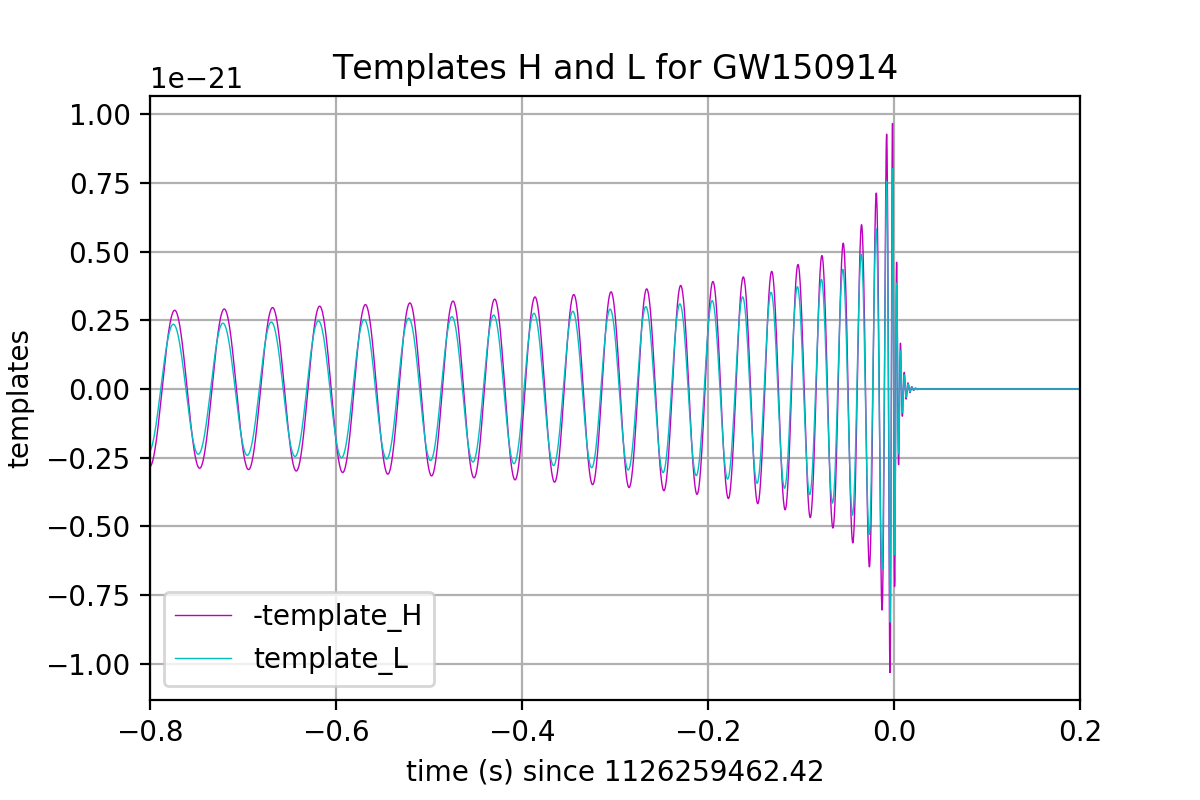}
\includegraphics[clip,width=0.48\textwidth]{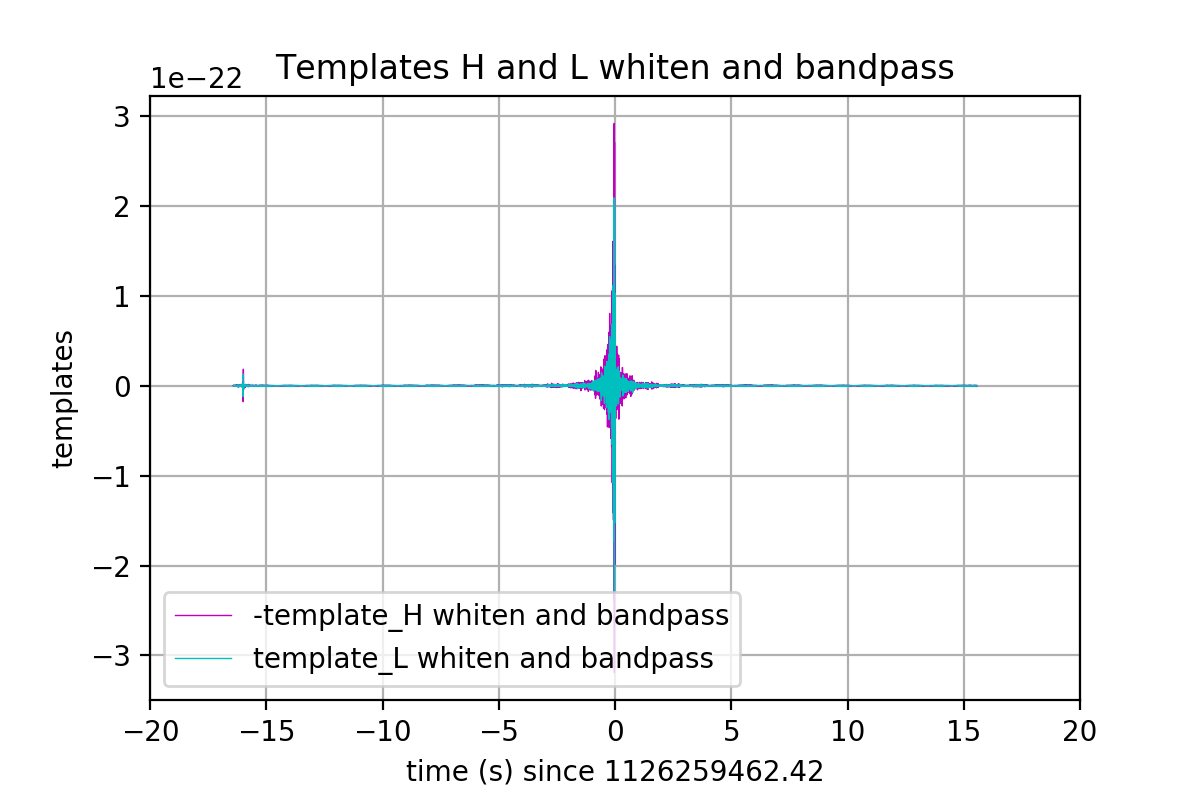}
\includegraphics[clip,width=0.48\textwidth]{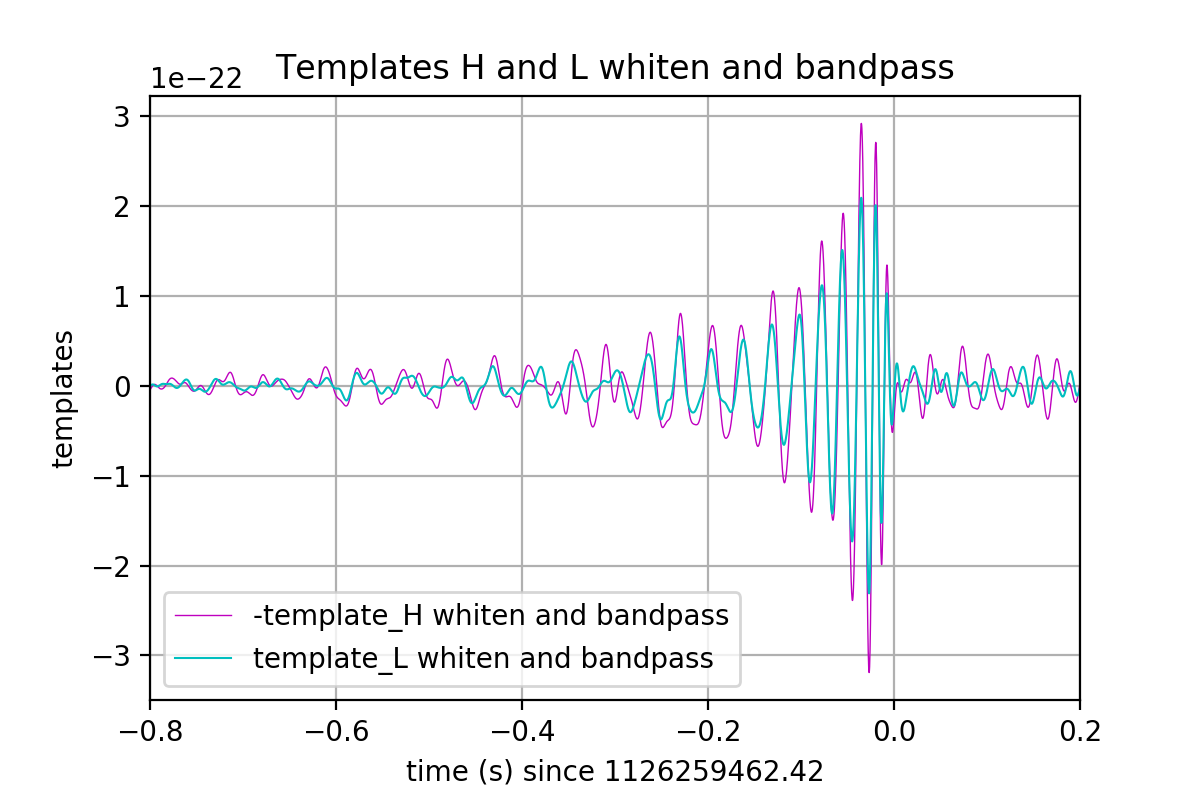}
\caption{
	Matched templates for Hanford and Livingston GW150914 event.
	In top left graph, the full time extend of the original templates; 
	while on the
	top right, the detail of 1s length around the time of the event.
	In the bottom row, the corresponding graphs after applying the
	whitening and bandpass of 35-350Hz filter. 
}
\label{fig:templates-1}
\end{figure}
One can see in the graphs of figure \ref{fig:templates-1} that the filtering
procedure suggested by LIGO, severely changes the shape of the matched templates;
which are supposed to give a close representation of the expected physical
signal hidden in the observed data.
In particular, in the lower right graph, it is noted that by applying
these filters, one can only use a very limited lapse of time in the
signal, on the order of 0.1second.
We will argue below against this limitation imposed by the filtering approach
suggested by the LIGO Collaboration.

\section{The new filtering scheme without whitening}\label{sec:newfiltering}

A comparison of the U shape shown in figure \ref{fig:ASD-raw} 
and the flat shape shown in \ref{fig:whiten} indicates that
if there were a physical signal with frequencies below that one
at minimum of the ASD; then they will be severely attenuated
by the whitening procedure. 
We present here a new approach to the initial filtering in order
to circumvent this effect.

\subsection{Initial bandpass filter}

It is sensible to apply an initial bandpass filter in order to concentrate
on the main frequency band that might contain physically interesting signals.
For this reason we choose a very wide band that goes from 22Hz to 1024Hz.
However, instead of the infinite impulse response (IIR) filters
used in the public LIGO python scripts,
we choose to apply finite impulse response (FIR) filters;
which are supposed to have safer phase behavior.

\begin{figure}[H]
\centering
\includegraphics[clip,width=0.49\textwidth]{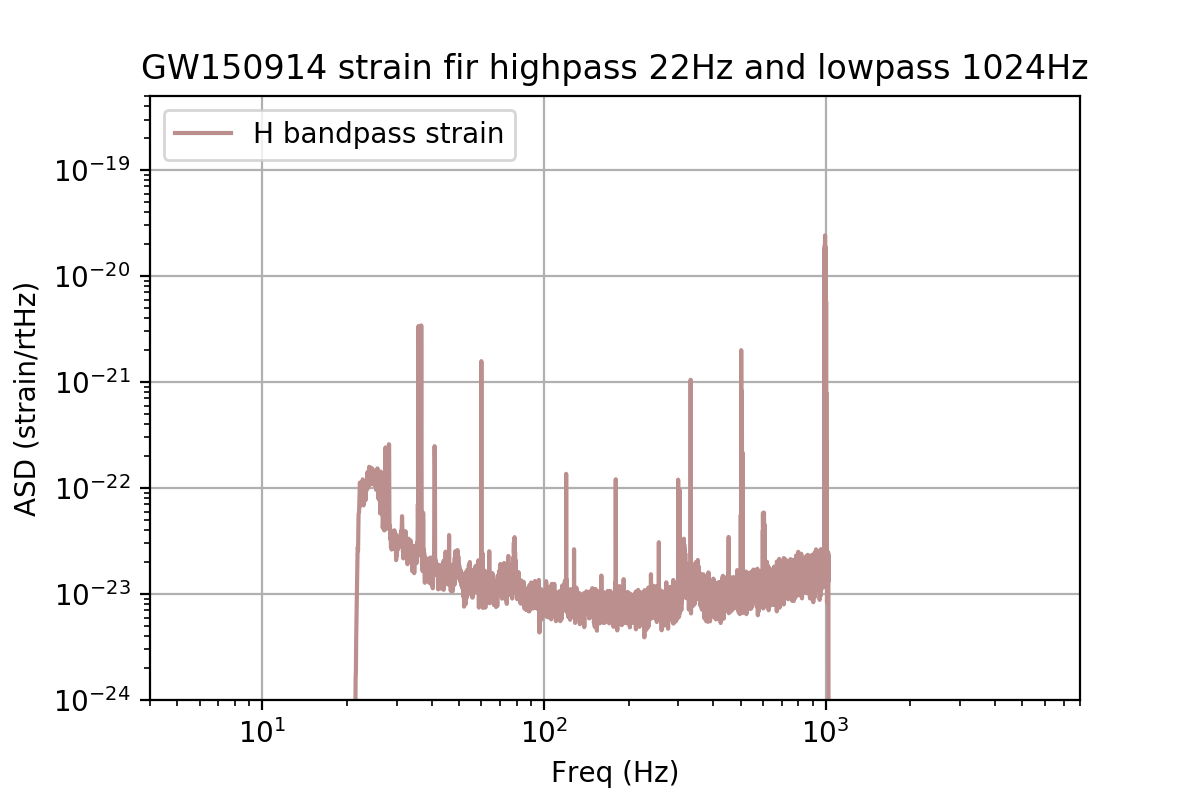}
\includegraphics[clip,width=0.49\textwidth]{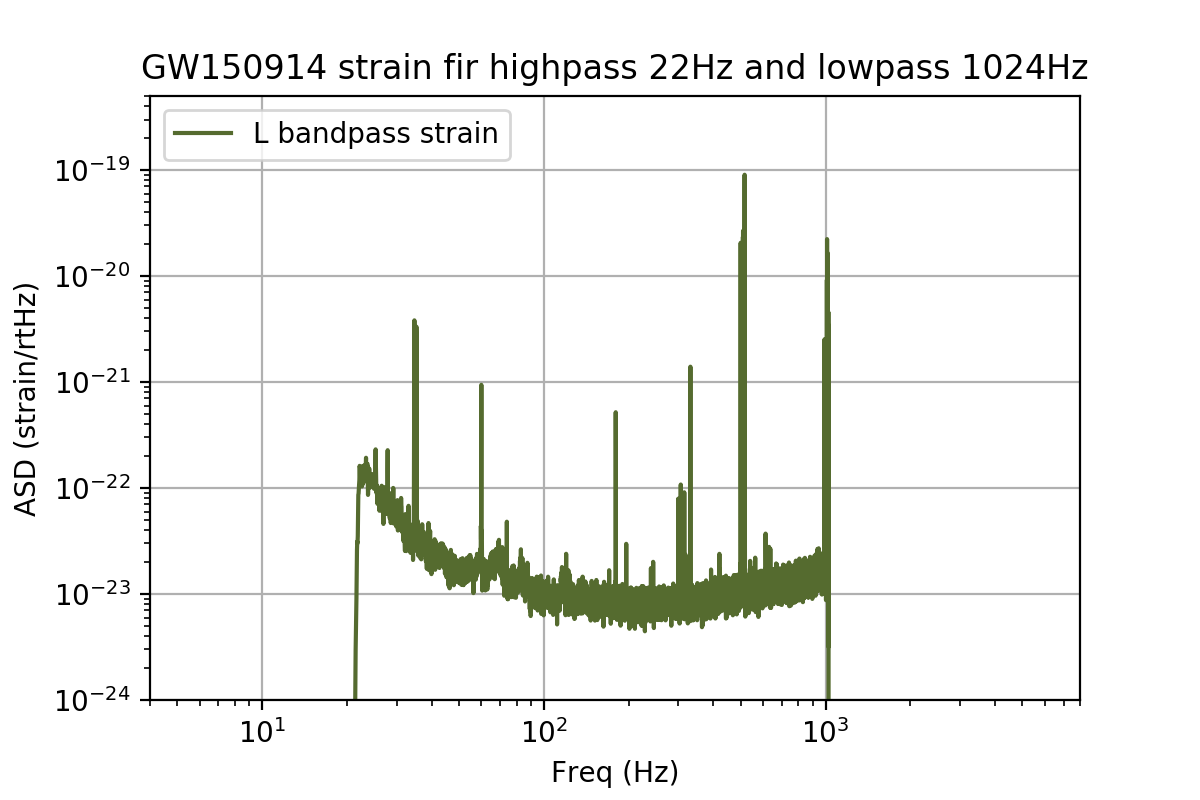}
\caption{Amplitude spectral density of the strains after bandpass FIR filter
	at frequencies 22-1024Hz; Hanford on the left and Livingston on the right.
}
\label{fig:osvfilt-hipa-lopa}
\end{figure}

Since we are interested in an interval of 256s and the FIR bandpass filters
might include boundary effects, we start with a 288s interval centered
at the time of the event, and after applying the FIR bandpass filter,
we crop it to an interval of 256s centered at the event time,
whose resulting ASD are shown in figure \ref{fig:osvfilt-hipa-lopa}.
Comparing with the graphs in figure \ref{fig:ASD-raw} one can see the sharp
attenuations at the chosen boundary frequencies.
One can still observed the narrow bands of intrinsic detector noise
at well defined frequencies; that we handle next.

\subsection{Filtering intrinsic detector frequencies}
A common way to deal with signals contained in complex noisy data is to
apply the whitening filters; that had been using the LIGO Collaboration.
Instead of this, we use stopband FIR filters for each of the
well defined narrow frequencies generated by each of the detectors.

\begin{figure}[H]
\centering
\includegraphics[clip,width=0.49\textwidth]{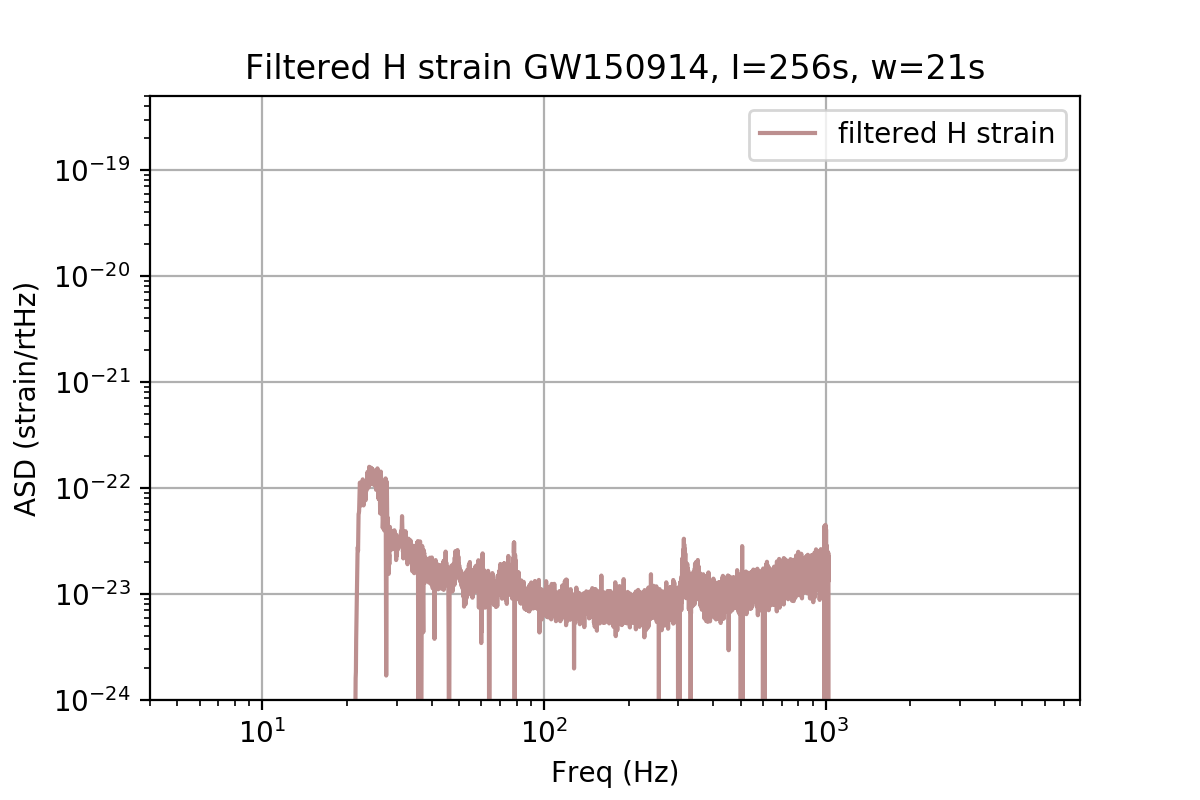}
\includegraphics[clip,width=0.49\textwidth]{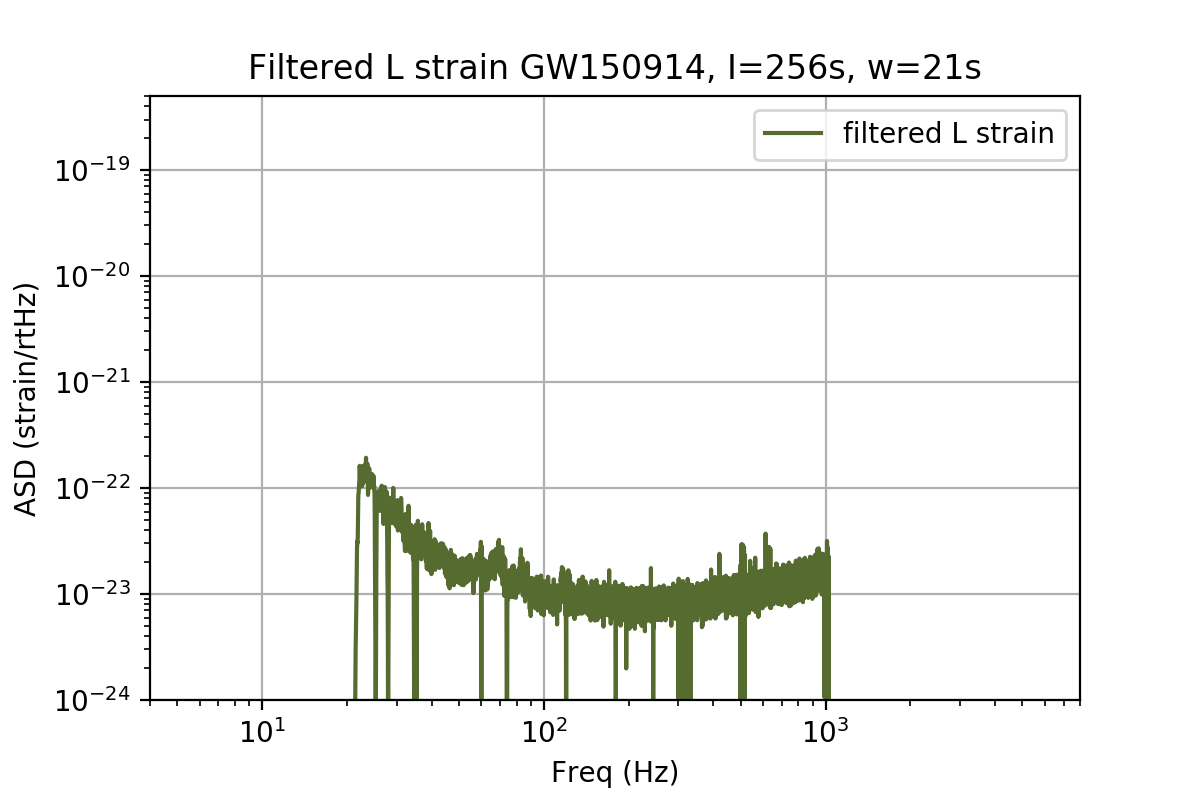}
\caption{Amplitude spectral density of the strains after bandpass and stopband FIR filter;
 Hanford on the left and Livingston on the right.
 We use the same axis limits as in previous graphs in order to facilitate the comparison.
}
\label{fig:osvfilt-hipa-lopa+stopbands}
\end{figure}

The first thing to do is to identify precisely the value of the frequency
and width of every intrinsic instrumental excitation introduced by each detector in the strain.
After this we apply the stopband FIR filter to each strain.
Again, to avoid boundary effects, we apply the filter to an interval of length 272s
which was obtained from the bandpass strain of 288s by clipping the extremes.
Then, we trim again the interval to the desired length of 256s.
The result is shown in figure \ref{fig:osvfilt-hipa-lopa+stopbands}.

It can be seen that the resulting ASD behavior is fairly flat
except for the initial curved behavior at low frequencies;
which it could be related to colored noise.
The idea behind the decision of not filtering the low frequencies at this stage
is that we would like to allow for possible interesting
physical signals at this initial portion of the spectrum.
Below, we do find such low frequency signals.

The strategy is then to avoid making changes to the astrophysical signal
in the pre-processing filtering stage.
This is completely different from the approach used up to now 
in LIGO/Virgo Collaboration 
articles\cite{Abbott:2016blz,TheLIGOScientific:2016wfe,
	TheLIGOScientific:2016qqj,TheLIGOScientific:2016uux}.

\subsection{Phase behavior after filtering}
\begin{figure}[H]
\centering
\includegraphics[clip,width=0.49\textwidth]{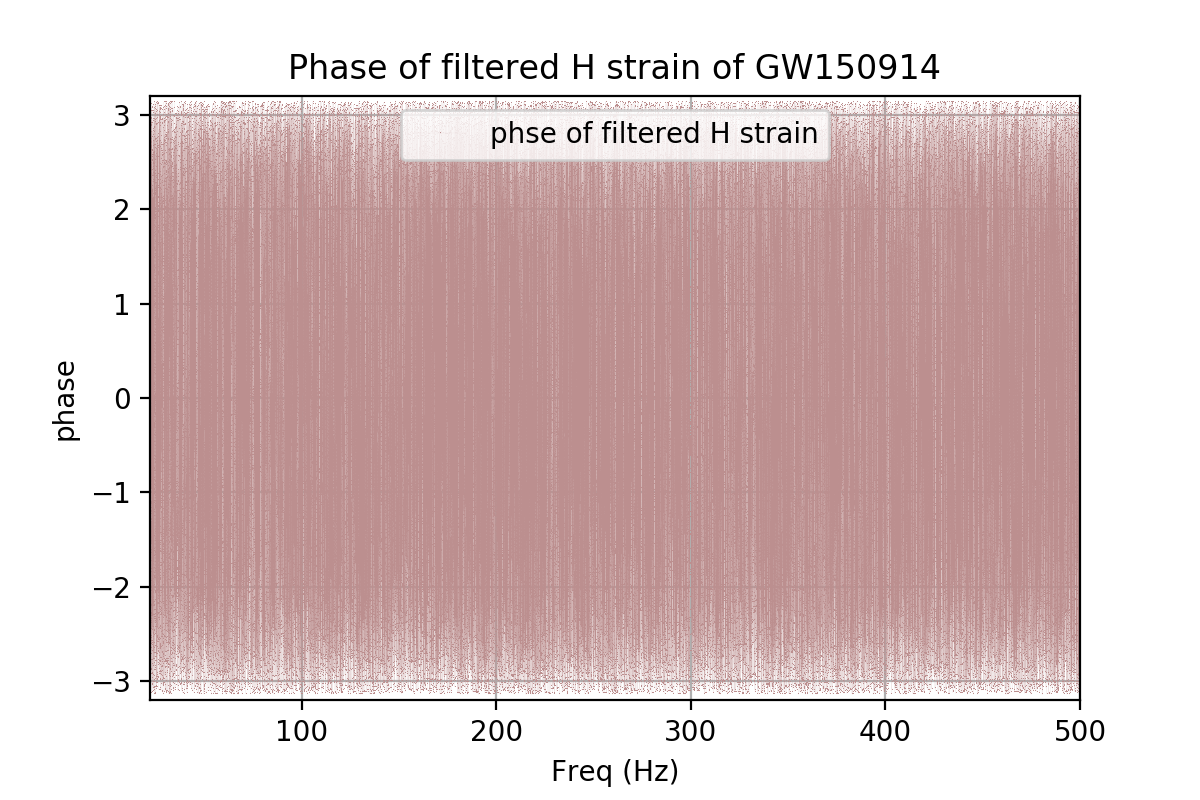}
\includegraphics[clip,width=0.49\textwidth]{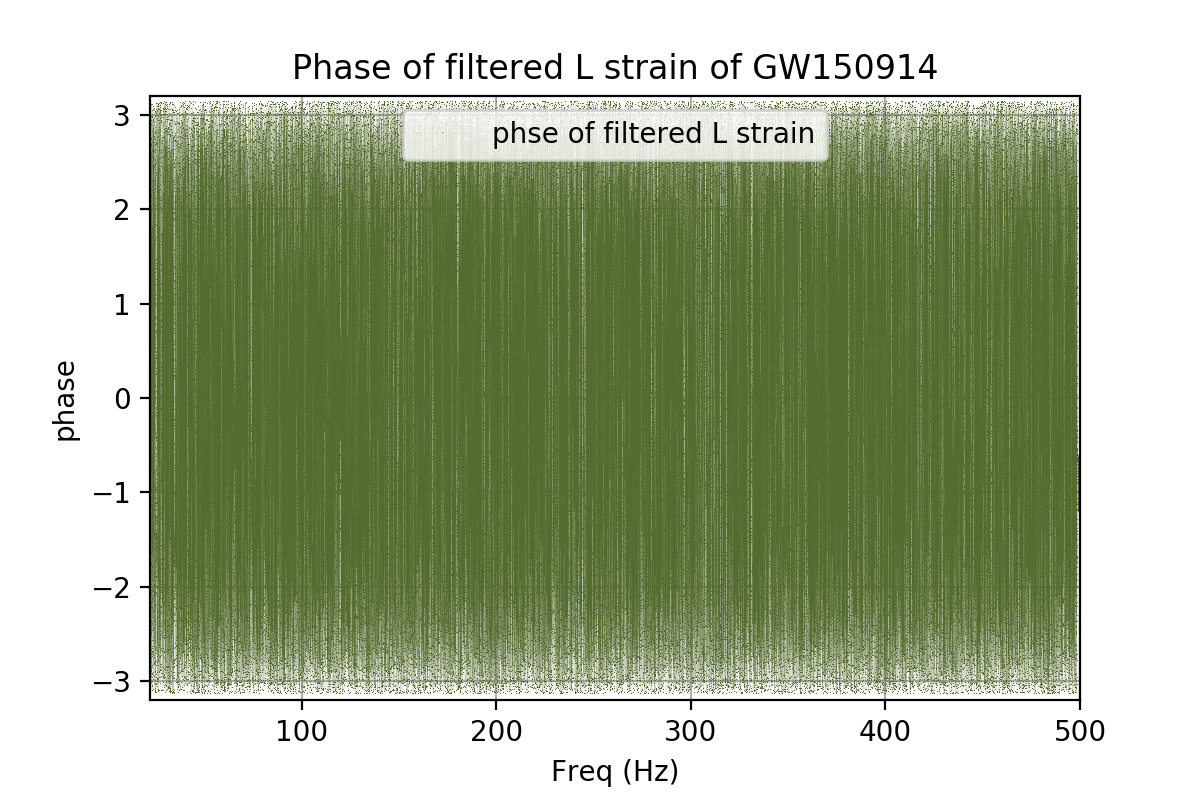}
\caption{Phase graphs of strains after new first FIR filtering round;
	Hanford on the left and Livingston on the right.
}
\label{fig:fase-osvfilt}
\end{figure}

In figure \ref{fig:fase-osvfilt} we show how the phases are distributed
across the frequency range of 22 to 500Hz, 
on the 256s strain of both detectors,
after the first set of FIR filters, bandpass and stopband, have been
applied. 
It can be seen that phases are evenly distributed in the frequency range of interest;
which it must be compared with the corresponding phase study after applying
the first LIGO filtering procedures to the strains, shown in figure \ref{fig:phase-filtered}.

This is a strong indication that we were able to suppress the initial phase
correlation in the raw data; and also that this correlation was due to the intrinsic
excitations of the detectors.

Of course an evenly distribution of phases is a good sign, since indicates
an almost Gaussian noise behavior.

\subsection{Time domain graphs after filtering}

Since the initial data is obtained in the time domain, we should
see what is the result of our filtering approach in it,
and also if we are able to obtain some new insight.

\begin{figure}[H]
\centering
\includegraphics[clip,width=0.75\textwidth]{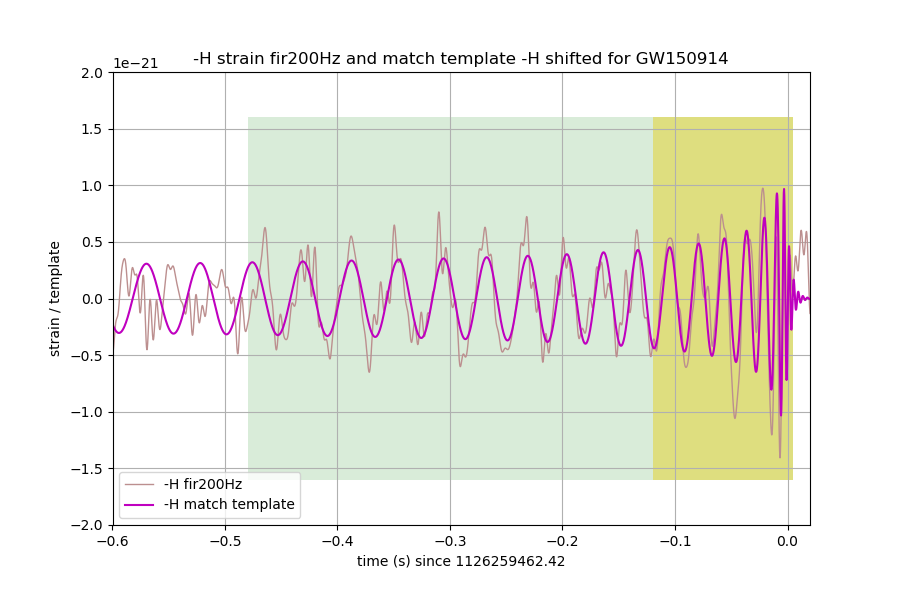}
\caption{
	Strain of Hanford detector after a lowpass 200Hz FIR-filter, along with
	its adjusted by LIGO script template,
	close to the time of the event. 
}
\label{fig:strain-H-time-domain+LIGOtemplates}
\end{figure}

\begin{figure}[H]
\centering
\includegraphics[clip,width=0.75\textwidth]{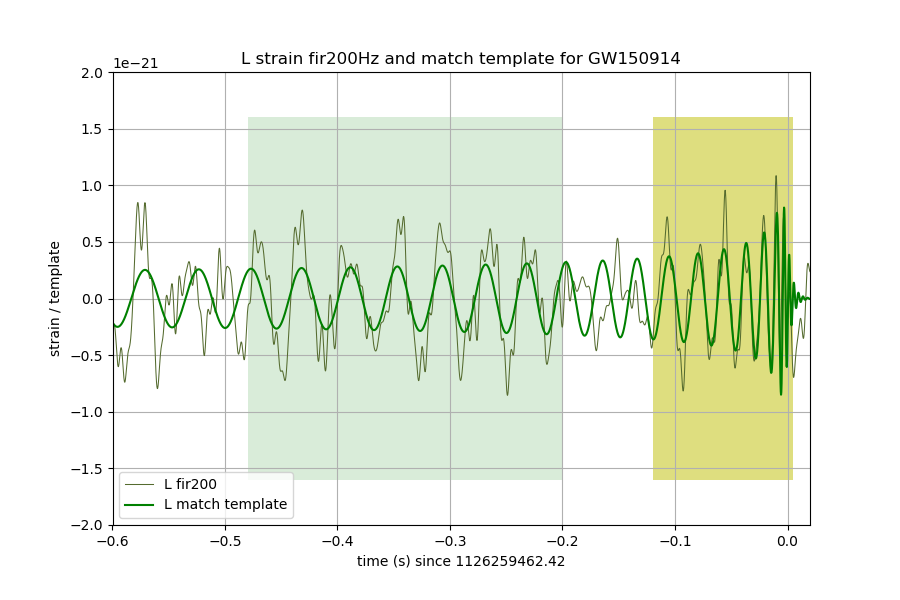}
\caption{
	Strain of Livingston detector after a lowpass 200Hz FIR-filter, along with
	its adjusted by LIGO script template,
	close to the time of the event.
}
\label{fig:strain-L-time-domain+LIGOtemplates}
\end{figure}

Figure \ref{fig:strain-H-time-domain+LIGOtemplates} and \ref{fig:strain-L-time-domain+LIGOtemplates}
presents respectively the graph
of the H and L strains along with the corresponding LIGO matched templates.
To avoid very high frequency contribution from the strain, that is not contained
in the template, we have applied for these graphs a lowpass FIR 200Hz filter.
It can be seen in the figure \ref{fig:strain-H-time-domain+LIGOtemplates} that
there is a remarkable coincidence in frequency, phase and amplitude of
the strain and the template in the time interval that goes from about -0.5s to -0.1s
before the time of the event; that we have marked with a light green color rectangle.
The light yellow band that includes the time of the event, and extends for about
0.1s, is to indicate approximately the lapse of time that is allowed by the
whitening LIGO procedure.
This was shown in the lower right graph of figure \ref{fig:templates-1} above.
Observing the Livingston graph of figure \ref{fig:strain-L-time-domain+LIGOtemplates}
one can see that there is a very good agreement in frequency, phase and amplitude of
the strain and the template in the time lapse that goes from about -0.5s to -0.2s
before the time of the event; that we have marked with a light green color rectangle.
There is a short period of time from about -0.2 to -0.1s where the detector
could not record properly the signal.

\begin{figure}[H]
\centering
\includegraphics[clip,width=0.75\textwidth]{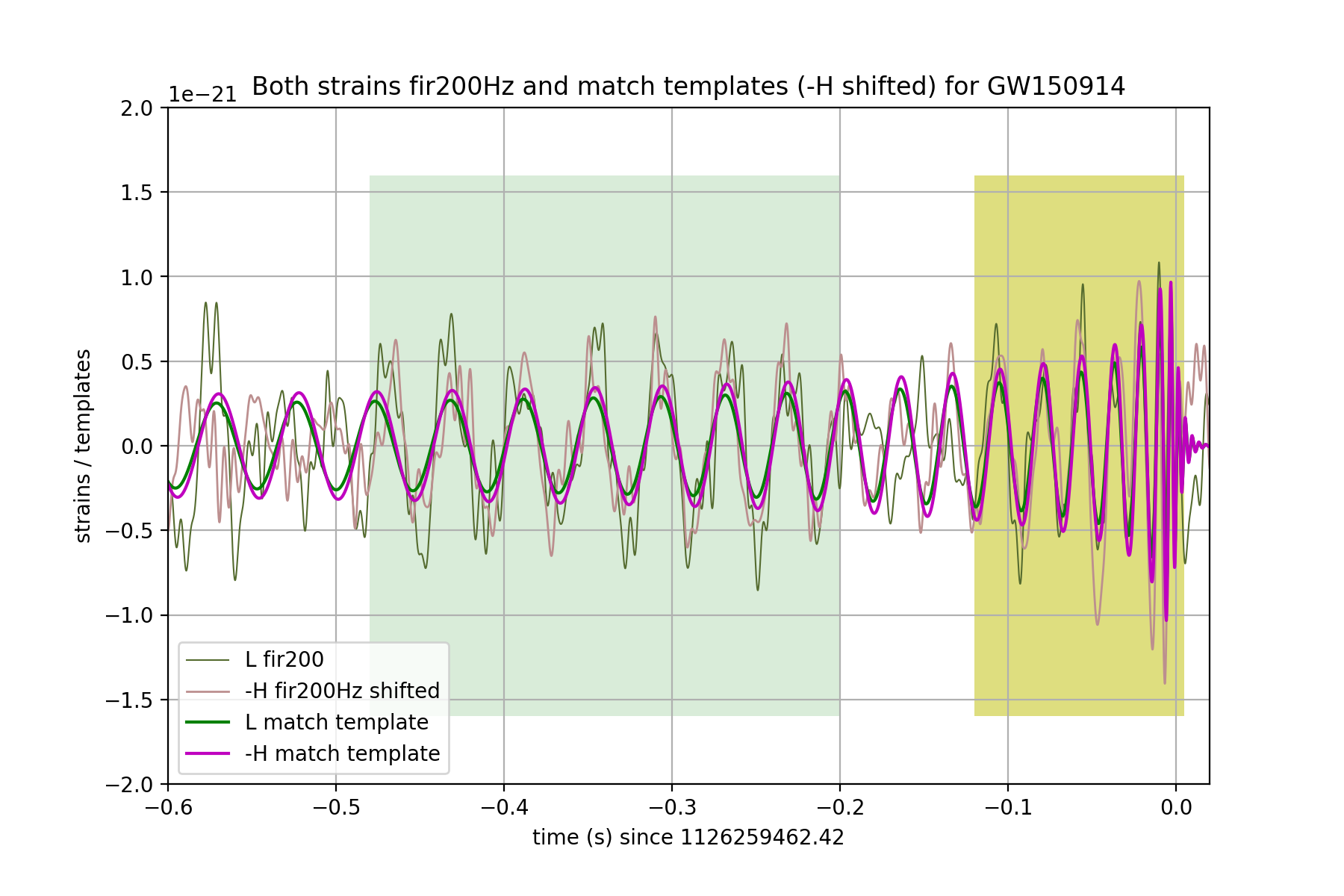}
\caption{
	Strain of both detectors after a lowpass 200Hz FIR-filter, along with
	their adjusted by LIGO script templates,
	close to the time of the event.
	}
\label{fig:strain-H+L-time-domain+LIGOtemplates}
\end{figure}

The fact that the same type of indication appears in both detectors
invites to present the two sets of data in the same graph,
which appears in figure \ref{fig:strain-H+L-time-domain+LIGOtemplates}.
In figure \ref{fig:strain-H+L-time-domain+LIGOtemplates} the H strain has been inverted
an appropriately shifted, in order to agree in phase at the high amplitude part of the signal
with the L strain.

It can be noticed that even if one considers only the light green region in which
both detectors show independently a match of the strain with the
corresponding templates, 
we find several coincidences:
the shown interval is very close to the event time,
at each detector the template matches the strain up to about -0.5s,
in phase, amplitude and frequency
and very importantly both detectors synchronize in the shown interval.
This is compelling evidence that the lapse of time
up to about -0.5 seconds before the time of the event
contains physically interesting signals.
Our findings can be contrasted with the claims in LIGO publications\cite{TheLIGOScientific:2016zmo}
in which they recognize up to 0.2s of signal.

This discovery shows precisely the benefit to use filtering techniques
that avoid to change the expected nature of the astrophysical signal.

\subsection{Spectrograms of the filtered signals}

\begin{figure}[H]
\centering
\includegraphics[clip,width=0.32\textwidth]{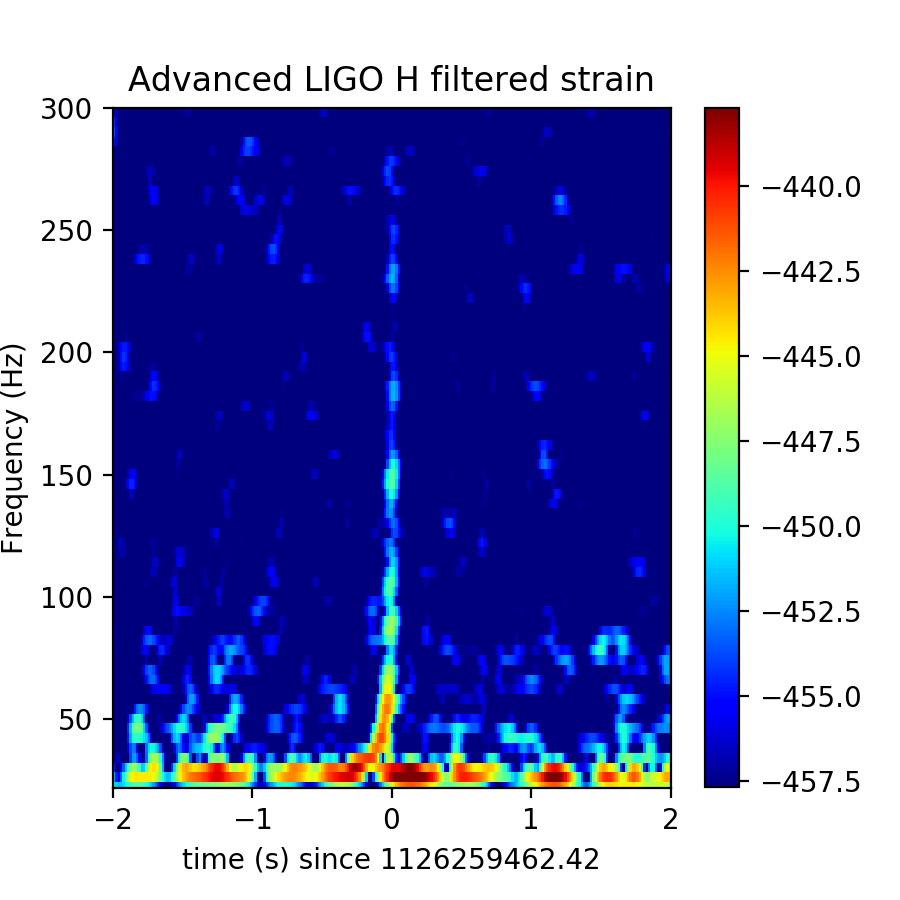}
\includegraphics[clip,width=0.32\textwidth]{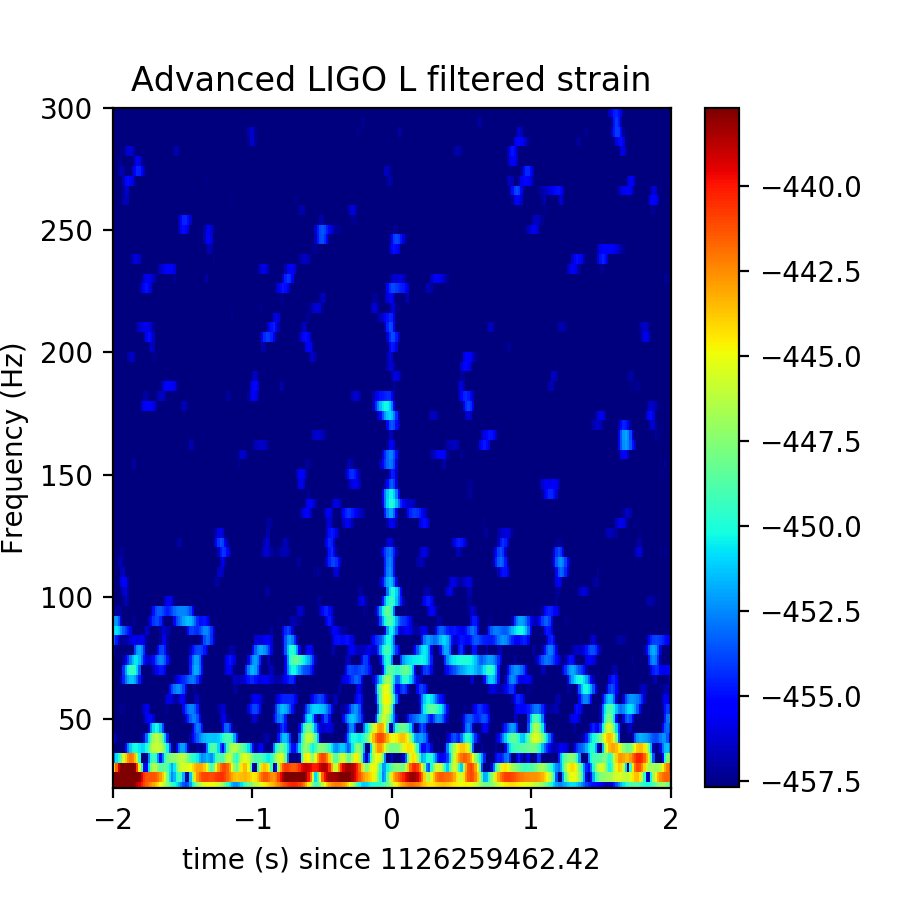}
\includegraphics[clip,width=0.32\textwidth]{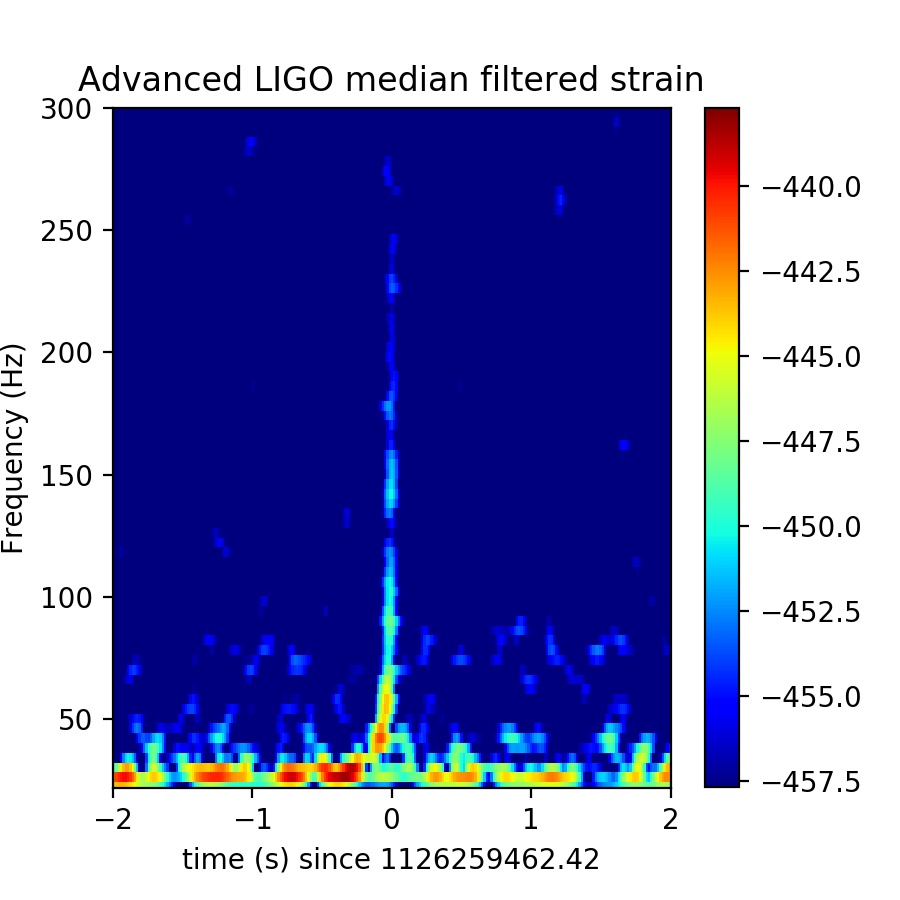}
\caption{Spectrograms of filtered data from Hanford on the left,
	and Livingston at the center; $\pm$2 seconds around the time of event GW150914,
	in the range from 22 to 300Hz.
We have also included on the right the corresponding spectrogram for the median;
where the H strain has been inverted and shifted.
}
\label{fig:spectrogrmas-filtered}
\end{figure}

In figure \ref{fig:spectrogrmas-filtered}
we present the graphs corresponding to the spectrograms after applying our filters
to the data, where one can see that the rising in frequency signals
reaches more than 250Hz in the Hanford data, and more than 200Hz in the
Livingston one.
Due to the chosen directions of the arms of the LIGO detectors,
Hanford and Livingston observatories will be sensible to almost the same component
of the polarization of the signal. This is the reason why both
matched templates are so similar.
Because of this one is tempted also to look at the median of
both strains, after the corresponding inversion and time shift
has been applied to one of the data; since in particular
by doing so one expects to obtain better reduction of the noise.
For this reason we have also included in the last graph of figure \ref{fig:spectrogrmas-filtered}
the corresponding spectrogram for the median.
There one can see that we have gain some noise reduction and also that
there seems to be some coincidence on the data
at low frequencies, within the last second before the time of the event;
which reinforces our claims observed in the time domain graphs.

\section{Final comments}\label{sec:final}

Let us summarize what we have presented so far.
In section \ref{sec:preliminar} we have reviewed
the basic characteristics of the data of the GW150914 event,
and showed the limitations in the pre-processing filtering techniques used
in the LIGO Collaboration studies; based on the whitening procedure,
and occasionally IIR bandpass filtering.

In order to circumvent the shortcoming of the whitening procedure
of too much attenuation for low frequencies, 
we have presented a new straight forward filtering approach based
on FIR filters, in section \ref{sec:newfiltering},
that performs and initial bandpass, and secondly
a careful stopband filter; that yields a minimum touched
strain which respects the possibility of low frequency
signals.

When observing the effect of our filter on the phase diagrams in figure \ref{fig:fase-osvfilt},
we deduce that our approach handles successfully the initial correlations
of phases shown in the raw data graphs, presented in section \ref{sec:preliminar}.
In this way we also avoid the awkward phase behavior after applying the
LIGO filtering techniques, shown in figure \ref{fig:phase-filtered}.

We have given evidence that, very close to the time
of the GW150914 event,
in a lapse of time of few tenths of a second:
there is coincidence of frequency, amplitude and phase,
between the detectors and with their respective templates used by the LIGO
Collaboration.
It is highly improbable that this concomitance should be attributed
just to chance. 
Therefore, there is more signal to be studied which is
encoded in the data of the GW150914 event.

This article is devoted to the presentation of a new approach
for the pre-processing of the data;
so that we intend to study in detail in separate works
the strain we have obtained in this way.
For this reason we have not quantified the coincidence
of strains at both detectors near the time of the event;
we do plan to present this with a new technique to
look for similar signals in a pair of detectors,
in a separate work.
We also intend to apply the filtering approach presented here
to all the available gravitational wave data of other events.

We suggest that many of previous analysis of the data,
should be carried out after applying the type of filtering we have
presented here; instead of the usual whitening pre-processing approach.


\subsection*{Acknowledgments}

We are very grateful to the LIGO/Virgo Collaboration for making available the
data and the python scripts on data analysis
at \href{https://www.gw-openscience.org/}{https://www.gw-openscience.org/}.

We thank Emanuel Gallo for a careful reading of the manuscript and for suggestions.

We acknowledge support from CONICET, SeCyT-UNC and Foncyt.



\end{document}